\documentclass[pre,aps,twocolumn]{revtex4-1}
\usepackage{graphicx}
\usepackage{amsmath}
\usepackage{indentfirst}

\DeclareMathAlphabet{\mathitbf}{OML}{cmm}{b}{it}

\newcommand{\sFrac}[2]{{\textstyle\frac{#1}{#2}}}

\newcommand{\rv}{\mathitbf r}

\newcommand{\calBold}[1]{\mbox{\boldmath${\cal #1}$}}

\newcommand{\Hes}{\calBold{H}}

\let \= \equiv \let\*\cdot \let\~\widetilde \let\^\widehat \let\-\overline

\begin{document}
\title{Stochastic Approach to Plasticity and Yield in Amorphous Solids}
\author{H.G.E. Hentschel$^{1,2}$, Prabhat K. Jaiswal$^1$, Itamar Procaccia$^1$ 
and Srikanth Sastry$^3$}
\affiliation{$^1$Department of Chemical Physics, The Weizmann Institute of
Science, Rehovot 76100, Israel
\\$^2$Department of Physics, Emory University, Atlanta, Georgia\\
$^3$ Jawaharlal Nehru Centre for Advanced Scientific Research, Jakkur Campus, 
Bangalore 560064, India}
%%%%%%%%%%%%%%%%%%%%%%%%%%%%%%%%%%
\begin{abstract}
We focus on the probability distribution function (pdf) $P(\Delta \gamma; 
\gamma)$ where $\Delta \gamma$ are the {\em measured} strain intervals between 
plastic events in an athermal strained amorphous solids, and $\gamma$ measures 
the accumulated strain. The tail of this distribution as $\Delta \gamma\to 0$ 
(in the thermodynamic limit) scales like $\Delta \gamma^\eta$. The exponent 
$\eta$ is related via scaling relations to the tail of the pdf of the 
eigenvalues of the {\em plastic modes} of the Hessian matrix $P(\lambda)$ which 
scales like $\lambda^\theta$, $\eta=(\theta-1)/2$. The numerical values of 
$\eta$ or $\theta$ can be determined easily in the unstrained material and in 
the yielded state of plastic flow. Special care is called for in the 
determination of these exponents between these states as $\gamma$ increases. 
Determining the $\gamma$ dependence of the pdf $P(\Delta \gamma; \gamma)$ can 
shed important light on plasticity and yield.  We conclude that the pdf's of 
both $\Delta \gamma$ and $\lambda$ are not continuous functions of $\gamma$. In 
slowly quenched amorphous solids they undergo two discontinuous transitions, 
first at $\gamma=0^+$ and then at the yield point $\gamma=\gamma_{_{\rm Y}}$ to 
plastic flow. In quickly quenched amorphous solids the second transition is 
smeared out due to the non existing stress peak before yield. The nature of 
these transitions and scaling relations
with the system size dependence of $\langle \Delta \gamma\rangle$ are discussed.
\end{abstract}
\maketitle
\section{Introduction}
\label{intro}

The response to external strain in amorphous solids as the strain $\gamma$ is 
increased in an athermal quasistatic manner (AQS) does not result in a smooth 
differentiable stress-strain relationship. Rather, it results in a sequence of 
reversible strain increases $\Delta \gamma_1, \Delta \gamma_2, \Delta \gamma_3 
\cdots$ followed in each case by an irreversible plastic event in which a local, 
sub-extensive or extensive group of particles dissipate energy and the system 
then falls into a new inherent state. After each such event $\Delta \gamma_i$ 
the stress has changed by $\Delta \sigma_i$. An ensemble of sufficiently many 
independent realizations of this
process at any value of $\gamma$ will then give rise to a probability 
distribution function (pdf) $P(\Delta \gamma;\gamma)$. The properties of this 
pdf, and in particular of the scaling form of the tail of this
pdf as $\Delta \gamma\to 0$ were the subject of focused recent research. The 
tail of the distribution is
relevant for the thermodynamic limit since the intervals $\Delta \gamma_i$ 
shrink when the system size
increases as shown below, Eq.~(\ref{meangam}).

In a pioneering paper in 2010 \cite{10KLP} it was shown that for $\gamma=0$, 
i.e. after a quench from the liquid state to the amorphous state at $T=0$ the 
tail
of the pdf appears to scale like
\begin{equation}
\lim_{\Delta \gamma\to 0} P(\Delta \gamma,\gamma=0) \sim (\Delta \gamma)^\eta\ . 
\label{defeta}
\end{equation}
The exponent $\eta$ was measured there to be $\eta\approx 0.6$ in 2-dimensions 
without any claim for universality; in particular this exponent may depend on 
the quench rate from the liquid to the solid. Besides the presence of a scaling 
tail it was shown that the whole pdf agrees extremely well with the Weibull 
distribution, giving rise to an interesting scaling relation to the {\em system 
size} dependence of the mean value of $\Delta \gamma$ at $\gamma=0$,
\begin{equation}
\langle \Delta \gamma\rangle \sim N^{\beta} \ , \quad \beta<0 \ . 
\label{meangam}
\end{equation}
The scaling relation follows from the theory of extreme value statistics, and 
with the Weibull form it reads \cite{10KLP}
\begin{equation}
\beta = -\frac{1}{1+\eta} \ , \quad \eta=-\left(1+\frac{1}{\beta}\right) \ . \label{stat}
\end{equation}
Indeed, the exponent $\beta$ was found to be in perfect consistency with 
$\eta\approx 0.6$, i.e.
$\beta\approx -0.62$. The method of measurement employed many independent 
realizations of quenched systems at $\gamma=0$. It is important to stress that 
for large systems the mean value $\langle \Delta \gamma \rangle$ tends to 
decrease with system size according to Eq.~(\ref{meangam}). The scaling 
presented in Eq.~(\ref{defeta})
refers always to values of $\Delta \gamma\ll  \langle \Delta \gamma \rangle$.

Other statistically stable scaling relations can be obtained in the steady state 
plastic flow after
the plastic yield. Here we have many repeated events in stationary conditions, 
allowing us to measure carefully the statistics of energy drops $\Delta U_i$, 
stress drops $\Delta \sigma_i$ and strain intervals $\Delta \gamma_i$. It was 
found numerically in \cite{09LP} and then analytically in \cite{10KLP} that
\begin{equation}
\langle \Delta U \rangle =\bar\epsilon N^\alpha\ , \langle \Delta \sigma\rangle 
=\bar s N^\beta\ , \langle \Delta \gamma\rangle =N^\beta \ ,
\end{equation}
with $\alpha=1/3$ and $\beta=-2/3$ as exact universal results. On the other hand 
it was shown that
the relation (\ref{stat}) is no longer valid in the plastic flow state. In fact 
there $\eta=0$ and the
pdf has no relation to the Weibull distribution. This major difference was 
ascribed to the existence of
sub-extensive plastic events
in the plastic flow steady-state.  These sub-extensive events are correlated, 
destroying
the statistical independence of repeated realizations in the quenched $\gamma=0$ 
state which is
necessary for the scaling laws (\ref{stat}).

The changes in the statistical properties as they are reflected in the pdf of 
the strain intervals $\Delta \gamma$ raises an interesting challenge. Can one 
follow these changes to understand plasticity in amorphous solids, and in 
particular to
study the yielding transition to
the steady state plastic flow. This challenge was picked up in particular in 
Refs.~\cite{15LW,15LGRW}. The upshot of these studies is that there exists a 
continuous change, as a function of the external
strain $\gamma$, of the values of the scaling exponent, from $\gamma=0$ to the 
plastic flow regime. If correct this would supply an extremely rare occasion of 
a continuous change in a scaling exponent as a function
of a control parameter \cite{15RDD}. The aim of this paper is to study carefully 
this interesting proposition.
We end up offering a different point of view in which we stress two 
discontinuous transition, one at
$\gamma=0^+$ and the other at the yield point $\gamma=\gamma_{_{\rm Y}}$. It 
should be pointed out however
that in this paper we discuss only the measurable strain intervals between 
plastic events and the lowest
eigenvalues of the Hessian matrix that vanish during the approach to a plastic 
events. In Refs. \cite{15LW,15LGRW} there appear distributions of variables that 
are not readily measurable in strained
amorphous solids, and a full comparison of predictions is not always easy. 
Nevertheless our conclusion
that the yield is a discontinuous transition (at least in slowly quenched 
amorphous solids) is in accordance
with recent studies that have used very different methods of investigation 
\cite{15KB}.

In Sect. \ref{numerics} we present some numerical estimates of the exponent 
$\beta$ and of the pdf
$P(\Delta \gamma;\gamma)$ for a range of values of the external strain $\gamma$. 
To interpret properly
the numerical results we turn in Sect. \ref{theory} to theoretical 
consideration. These considerations
 are based on relating the scaling properties of $P(\Delta \gamma;\gamma)$ to 
the scaling properties of
 another pdf, $P(\lambda;\gamma)$, which describes the distribution of 
eigenvalues of the {\em plastic modes}
 of the Hessian matrix, precisely those eigenvalues that approach zero before a 
plastic event is taking place. The separation into Debye modes and plastic modes 
was justified in \cite{11HKLP,12KLP,15GKPP}, and see also \cite{15GL}. In 
Sect.~\ref{theory} we
present a Fokker-Planck equation for the development of $P(\lambda;\gamma)$ as a 
function of
increasing $\gamma$. While we cannot vouch for the exactness of this equation, 
we argue that the scaling
properties of this equation are robust and can be trusted to provide the correct 
scaling exponents for the
tails of $P(\lambda;\gamma)$ and $P(\Delta \gamma;\gamma)$. In 
Sect.~\ref{conclude} we offer a summary
and conclusions.

\section{Direct Numerical Results}
\label{numerics}

In this section we describe direct numerical measurements of the exponent 
$\beta$ and $\eta$. To generate
data we have employed a 50-50 binary mixture of point particles interacting via 
Lennard-Jones potential
in 2-dimensions. The parameters of the model can be found in Ref. 
\cite{15GJPSZ}. The system is quenched from a high temperature liquid to a 
target temperature $T=0.001$ in Lennard-Jones units. We use two quench rates. In 
the first set of simulations the quench is
``infinitely fast" in the sense that we use conjugate gradient energy 
minimization.  In the second we reduce the temperature at a rate of $10^{-5}$ in 
Lennard-Jones units (cf. Ref.~\cite{09LP}), and we refer to it below as the 
``slow" quench. The mechanical properties of the differently quenched systems 
differ. The slow quench shows
a distinct stress peak when $\gamma$ is increased, followed by a yield toward 
the steady state plastic flow. The fast quench results in a gradual increase in 
stress towards the plastic flow steady state without
a stress peak.

After preparing the system in athermal condition we follow the standard AQS 
protocol to strain the system
in simple shear strain. We carefully measured the strain intervals $\Delta 
\gamma_i$ that occurred between
plastic events. We back-tracked our simulations to increase the precision of 
this measurements (cf. Appendix of \cite{09LP}). Every such simulation was 
repeated using 1000-1500 freshly quenched independent realizations. We have 
measured the distribution of $\Delta \gamma_i$ and the mean value $\langle 
\Delta \gamma \rangle$ as a function of $N$ for different system sizes. While 
there is no ambiguity about the latter quantity at $\gamma=0$, for higher values 
of $\gamma$ we need to collect data within a bin of values of $\gamma$. We opted 
to do so in bins of size 0.01. Thus where we report below a value of $\gamma$, 
say $\gamma=0.02$ one needs to interpret that
as data collected from all events that occur between $\gamma=0.01$ and 
$\gamma=0.02$.

\begin{figure}
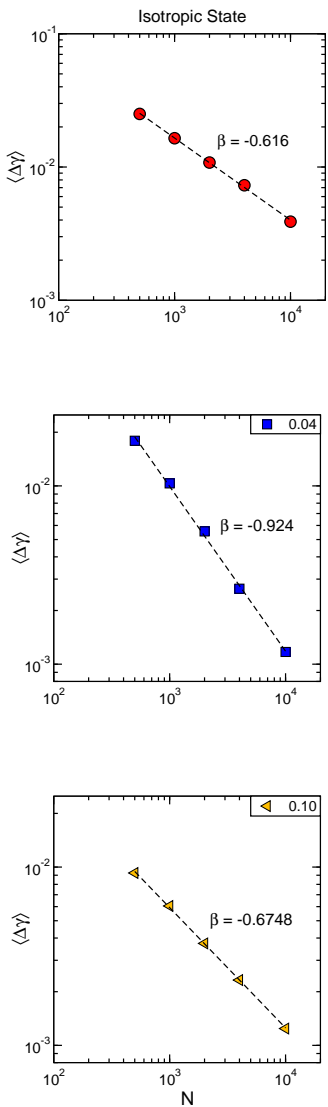

\includegraphics[scale = 0.26]{FPFig1a.eps}
\vskip 0.9 cm
\includegraphics[scale = 0.26]{FPFig1b.eps}
\vskip 0.9 cm
\includegraphics[scale = 0.26]{FPFig1c.eps}
\caption{The scaling law (\ref{meangam}) demonstrated for $\gamma=0$ (upper 
panel), $\gamma=0.04$ (middle panel)
and $\gamma=0.1$ (lower panel). We propose that in the range 
$0<\gamma<\gamma_{_{Y}}$ the value of
$\beta$ in the thermodynamic limit is $\beta=-1$. }.
\label{simbeta}
\end{figure}

\begin{figure}
\includegraphics[scale = 0.35]{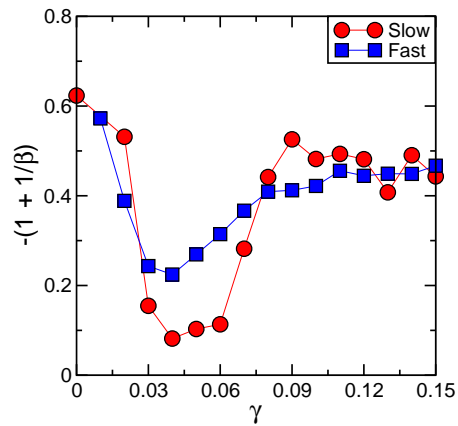}
\caption{A plot of $-(1+\frac{1}{\beta})$ obtained from the direct measurement 
of $\beta$ as a function
of $\gamma$. Results for the slow quench are shown in circles and for the fast 
quench in squares. We argue in this paper that these results suffer from severe 
finite size effects and in reality this figure should be
replaced by Fig.~\ref{final}}.
\label{mislead}
\end{figure}

In Fig.~\ref{simbeta} we show some representative results for the measurement of 
the exponent $\beta$.
In the upper panel the scaling law (\ref{meangam}) is demonstrated for 
$\gamma=0$, whereas in the lower
panels we show the result for $\gamma=0.04$ and $\gamma=0.1$. The first one is 
in agreement with
the previous estimate of $\beta\approx -0.62$. The last one is beyond the 
plastic yield and is in
agreement with the exact prediction $\beta=-2/3$. The middle panel is an example 
of the new numbers
obtained here for the first time for intermediate values of $\gamma$.
%%%%%%%%%%%%%%%%%%%%%%%%%%%%%%%%%%%%%%%%%%%%%%
An over-all impression of the resulting picture can be obtained {\em if we 
assume that the scaling
law (\ref{stat}) is obeyed throughout the range of $\gamma$.} We can then use 
the measurement of $\beta$ to
plot $-(1+\frac{1}{\beta})$ as a function of $\gamma$, see Fig.~\ref{mislead}.
If one believes that this represents the exponent $\eta$ according to 
(\ref{stat}), then this continuous variation as a function of $\gamma$ is somewhat 
reminiscent
of the dependence of this exponent on $\gamma$ as predicted
in Refs.~\cite{15LW,15LGRW} (This exponent is denoted as $\theta$ there).
In fact we will argue now that (i) the scaling law (\ref{stat}) is valid only as 
long as $\gamma<\gamma_{_{\rm Y}}$. (ii) The measured values of $\beta$ for 
$0<\gamma<\gamma_{_{\rm Y}}$ suffer from severe finite size and cross-over 
effects. In fact the analysis presented below indicates that $\eta=0$ and 
$\beta=-1$ for all the values of $\gamma$ in the range $0<\gamma<\gamma_{_{\rm 
Y}}$, with a transition to $\beta=-2/3$ after the
plastic yield transition. The transition is expected to be sharp in the case of 
the slow quench, but much
more smeared out in the case of the fast quench.

\begin{figure}
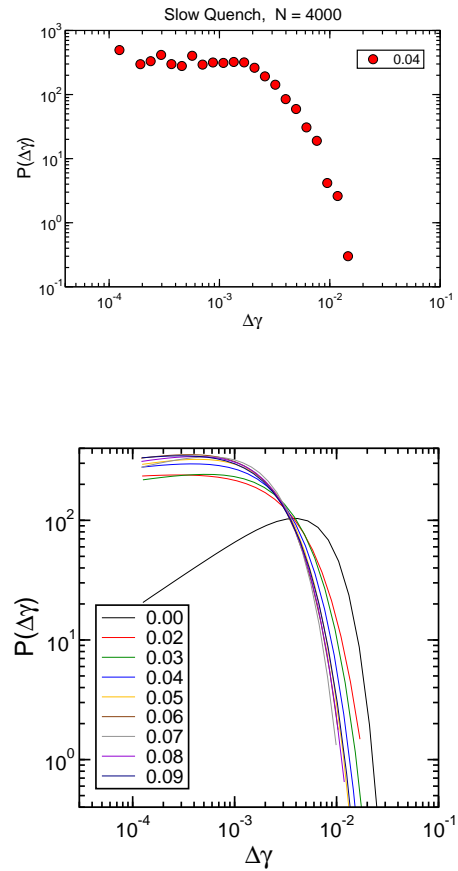

\includegraphics[scale = 0.25]{FPFig3a.eps}
\vskip 1.5 cm
\includegraphics[scale = 0.35]{FPFig3b.eps}
\caption{Results for the slow quench. Upper panel: an example of the raw data 
for the probability distribution functions $P(\Delta \gamma;\gamma)$ for 
$\gamma=0.04$. Lower panel: the fits to the pdf's $P(\Delta \gamma;\gamma)$ for 
a range of values of $\gamma$ from $\gamma=0$ to $\gamma=0.9$ (beyond the 
plastic yield). Only at $\gamma=0$ the
distribution is Weibull with $\eta\approx 0.6$. For all other values of $\gamma$ 
we find $\eta=0$. This
value is explained by the theory in Sect.~\ref{theory}.}
\label{pdfs}
\end{figure}

To see that the scaling law (\ref{stat}) fails for $\gamma>\gamma_{_{\rm Y}}$ it 
is sufficient to examine the pdf's $P(\Delta \gamma;\gamma)$. In Fig.~\ref{pdfs} 
we present these distribution functions for the slowly quenched amorphous solid 
for a range of $\gamma$ values
starting with $\gamma=0$ and ending with $\gamma=0.09$ which is already beyond 
the plastic yield
which for the present slowly quenched system is estimated at $\gamma_{_{\rm 
Y}}\approx 0.06$. It is
clear that for $\gamma=0$ the pdf has a perfect Weibull form with $\eta\approx 
0.6$ in agreement with
$\beta\approx -0.62$. But for all $\gamma>0$ the pdf has a flat tail as $\Delta 
\gamma\to 0$, indicating
a value of $\eta=0$. In the plastic flow state, at $\gamma>\gamma_{_{\rm Y}}$, 
where we measure $\beta=-2/3$,
the scaling law (\ref{stat}) is not obeyed. It was proposed in Ref.~\cite{10KLP} 
that this is due to the  sub-extensive plastic events that occur in the flowing 
state. The arguments leading to the scaling
law (\ref{stat}) fail under these conditions.
%%%%%%%%%%%%%%%%%%%%%%%%%%%%%%

\begin{figure}
\includegraphics[scale = 0.35]{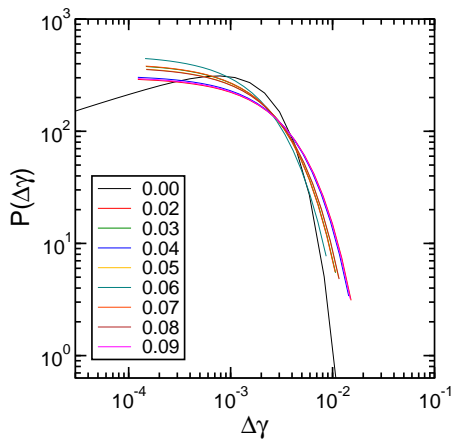}
\caption{Results for the fast quench: the probability distribution functions 
$P(\Delta \gamma;\gamma)$ for a range of values
of $\gamma$ from $\gamma=0$ to $\gamma=0.9$ (beyond the plastic yield). Only at 
$\gamma=0$ the
distribution is Weibull with $\eta\approx 0.3$. For all other values of $\gamma$ 
we find $\eta=0$. This
value is explained by the theory in Sect.~\ref{theory}.}
\label{fastq}
\end{figure}

The question is then whether the scaling law (\ref{stat}) holds in the range 
$\gamma<\gamma_{_{\rm Y}}$.
We will argue that the answer is in the affirmative. We read $\eta=0$ from the 
results of Fig.~(\ref{pdfs}) and we propose that the correct value of $\beta$ is 
$\beta=-1$, in agreement with the scaling law (\ref{stat}).
This is equally valid for the slow and the fast quench (see the results of the 
pdf's for the slow quench in
Fig.~\ref{fastq}).
The results shown in Fig.~\ref{mislead} should be read with this in mind; due to 
system size and
crossover effects the sharp transition at $\gamma=0^+$ is smeared out, and so is 
the transition near
$\gamma_{_{\rm Y}}$. Clearly, in the case of the fast quench the transitions are 
even more smeared out. We propose that in the case of slow quench, in the first 
transition $\eta$ tries to reach the value $0$ and in the second transition
the scaling law (\ref{stat}) breaks down and $\beta$ settles on the value -2/3. 
The theoretical predictions
will be presented below in Fig.~\ref{final} after the considerations leading to 
these assertions become clear.

\section{Theoretical Considerations}
\label{theory}
\subsection{Relation to the Hessian matrix and its eigenvalues}

To understand the statistics of $\Delta \gamma$ we need to connect it to the 
statistics of the
eigenvalues of the Hessian matrix \cite{03GC,15BMPP}. The latter is crucial for 
the understanding of athermal plasticity.
 Each plastic instability is due to a saddle-node bifurcation in which the 
lowest eigenvalue $\lambda_i$ of the associated Hessian matrix  $\Hes_{ij}= 
\frac{\partial^2 U}{\partial {\rv}_{i}\partial {\rv}_{j}}$ of the potential 
energy $U$ softens and becomes zero at some value of $\gamma=\gamma_P$. The 
modes whose eigenvalues become zero at some value of the external strain are 
referred to as plastic modes \cite{11HKLP,12KLP,15GKPP}. One should be aware 
that every disordered solid has also Debye modes whose eigenvalues remain 
roughly independent of the external strains. For the plastic modes, because we 
are dealing with a saddle-node instability, we know (and see below for details) 
that $\Delta \gamma_i \sim \lambda_i^2$ and also that the energy barrier 
$\epsilon_i$ scales as $\epsilon_i \sim 
\lambda_i^3$. Analyzing the properties of an ensemble of  the lowest eigenvalues 
of the plastic modes of the Hessian matrix at fixed strain $\gamma$ will yield 
(in the thermodynamic limit) a distribution $P(\lambda ;\gamma )$. Knowing the 
tail of this pdf for
 $\lambda\to 0$,
  \begin{equation}
  \lim_{\lambda\to 0} P(\lambda ;\gamma ) \sim \lambda^\theta \ ,
  \end{equation}
  will allow the calculation of the tails of the distribution of the strain 
events $P(\Delta \gamma; \gamma)$ at fixed $\gamma$ and the tail of the 
associated  energy barrier distribution $P_\epsilon(\epsilon ;\gamma )$ at fixed 
$\gamma$. We reiterate that the distribution $P(\lambda ;\gamma )$ discussed 
here refers to
  the minimal eigenvalues that are exposed in the tail of the distribution as 
they tend to zero due to plastic
  processes. There may be other putative plastic modes that are not yet ready to 
express their potential
  instability and these are not taken into account here. A good example of these 
plastic modes are those discussed explicitly in Ref.~\cite{15GL}.

It is very important to keep separate in one's mind two sets of experiments. In 
the situation described above the strain $\gamma$ is fixed at some value and an 
ensemble of $\Delta \gamma_i$ measured. A second experiment involves ramping up 
$\gamma$ from zero and measuring consecutive reversible $\Delta \gamma_i$ 
followed by avalanching. The $n$th such event will occur at $\gamma = 
\Sigma_{i=1}^n \Delta \gamma_i$. In such an experiment the stochastic dynamics 
will prove to be very similar to particles drifting in a one dimensional flow 
together with added shot noise  (with the eigenvalue $\lambda$ representing the 
particle `coordinate' and $\gamma$ equivalent to time).
In the thermodynamic limit, when the distribution of eigenvalues of the Hessian 
becomes a density, we will employ a Fokker-Planck equation for $P(\lambda ; 
\gamma )$ which is similar in structure to a one dimensional flow for $\lambda$ 
with an adsorbing boundary at $\lambda = 0$ due to the associated saddle-node 
instability followed by reinjection of the of a new eigenvalue with a 
distribution $j_{in}(\lambda ,\gamma )$ in order to conserve probability. Its 
generic form will be
\begin{equation}
\label{fppdf}
\partial P/\partial \gamma = - \partial [v(\lambda , \gamma ) P] + D(\gamma) 
\partial^2 P/\partial \lambda^2 + j_{in}(\lambda ,\gamma ).
\end{equation}
For understanding the scaling properties of $P(\lambda ; \gamma )$ it is crucial 
to derive the scaling
form of $v(\lambda , \gamma )$ which we discuss next.
%%%%%%%%%%%%%%%%%%%%%%%%%%%%%%%%%%%%%%%%%%
\subsection{The rate of change of the eigenvalues and the resulting scaling 
relations}
The central result that will be used in the present context is an ``equation of 
motion" for the change in
the minimal eigenvalues of the Hessian matrix upon increasing strain. This 
equation was derived in Ref.~\cite{11HKLP}, and it turns out to be rather 
complex. Nevertheless, in the limit of $\lambda\to 0$ this equation can be 
simplified to exhibit only the most singular contribution, correct up to less 
singular terms:
\begin{equation}
\label{moo03}
\frac{d\lambda}{d\gamma} \simeq -\frac{a(\gamma )}{\lambda} +\text{less singular 
terms}.
\end{equation}
Together with the boundary condition that $\lambda$ vanishes at some strain 
value $\gamma_P$,
i.e. $\lambda(\gamma_P) = 0$, we integrate Eq.~(\ref{moo03}) to obtain
\begin{equation}
\label{eomSolution}
\lambda (\gamma ) = \sqrt{2a (\gamma_P) (\gamma_P - \gamma)}+\text{higher order 
terms}\ .
\end{equation}
We can also expand the projection of the potential energy in the direction
that is becoming unstable for $\gamma \to \gamma_P$
\begin{equation}
U = U_0 + \sFrac{1}{2}\lambda s^2 + \sFrac{1}{6}  b s^3 + {\cal O}(s^4)\ .
\end{equation}
This form implies that a saddle point exists at $s_\star = -\frac{2\lambda}{ 
b}$,
of magnitude
\begin{equation}
\label{ep}
\Delta E \equiv U(s_\star) - U_0 \simeq \frac{2\lambda^3}{3 b^2} =
\frac{4\sqrt{2}}{3}\sqrt{\frac{a^3}{ b^4}}\
(\gamma_P - \gamma)^\frac{3}{2}\ ,
\end{equation}
where we have used the solution Eq.~\ref{eomSolution} for the $\gamma$ 
dependence.
Eq.~(\ref{eomSolution}) and Eq.~(\ref{ep}) are the basic relations from which 
the scaling relations by which the strain increases and energy barriers are 
related to $\lambda$
\begin{eqnarray}
\label{scaling}
\Delta \gamma & \sim &\lambda^2 \label{rel} \\
\epsilon &\sim & \lambda^3 .
\end{eqnarray}
can be found. In particular Eq.~(\ref{rel}) leads to the obvious scaling 
relation
\begin{equation}
\eta =(\theta -1)/2 \ . \label{etatheta}
\end{equation}
We expect the scaling relation (\ref{etatheta}) to be valid for all values of 
$\gamma$ and for all
amorphous solids since it stems
from the saddle-node nature of the plastic events which is generic.
%%%%%%%%%%%%%%%%%%%%%%%%%%%%%%%%%%%%%%%%%%%%%%%%%%%%%%
\subsection{Consequences for the Fokker-Planck Equation }

 The upshot of the preceding discussion is that we can estimate the most 
singular form of the `speed' $v(\lambda;\gamma)$
 in Eq.~(\ref{fppdf}) as
 \begin{equation}
 v(\lambda;\gamma) \approx - a(\gamma)/\lambda + \text{less singular terms}\ . 
\label{drift}
 \end{equation}
We therefore rewrite Eq.~(\ref{fppdf}) in the form
 \begin{equation}
\label{fokkerplanck2}
\partial P/\partial \gamma \approx a(\gamma) \partial [ P/\lambda ]/\partial 
\lambda + D(\gamma)\partial^2 P/\partial \lambda^2 + j_{in}(\lambda,\gamma ) \,
\end{equation}
with a probability flux
\begin{equation}
J(\lambda ,\gamma ) =  - a(\gamma) P/\lambda - D(\gamma )\partial P/\partial 
\lambda \ . \label{J}
 \end{equation}
 The diffusive term in the Fokker-Planck equation results from the ``less 
singular terms" in Eq.~(\ref{drift}).
 Note that we should seek solutions of this equation with an adsorbing boundary 
condition at $\lambda = 0$, leading to
 \begin{equation}
\label{pad}
P(\lambda=0, \gamma ) = 0 ,
\end{equation}
but with a finite flux for any value of $\gamma$. We stress that we have added a 
non-singular (in $\lambda$) diffusion term to the drift term in the Fokker 
Planck equation. The reason is that the theory leading to
Eq.~(\ref{drift}) does not indicate the existence of any other singular term  in 
the ``noise" on top of
the leading,  explicit $1/\lambda$ term in Eq.~(\ref{drift}). The most singular 
contribution is already there.
For the same reason also the injection term $j_{in}(\lambda,\gamma )$ is taken 
as non-singular. 

Accepting the fact that for any value of $\gamma>0$ the flux must be finite, a 
direct consequence
of the form of Eq.~(\ref{J}) is that $P(\lambda; \gamma)$ {\em must} start 
linearly in $\lambda$ in
order to have a finite flux which is neither zero nor infinity:
\begin{equation}
\lim_{\lambda\to 0} P(\lambda; \gamma) \sim\lambda \ . \label{lamtozero}
\end{equation}
{\bf Using the scaling relation (\ref{etatheta}) we conclude that $\eta=0$ for 
any value of
$\gamma>0$. This result rationalizes entirely the data shown in 
Fig.~\ref{pdfs}.}

\begin{figure}
\hskip -0.5 cm
\includegraphics[scale = 0.4]{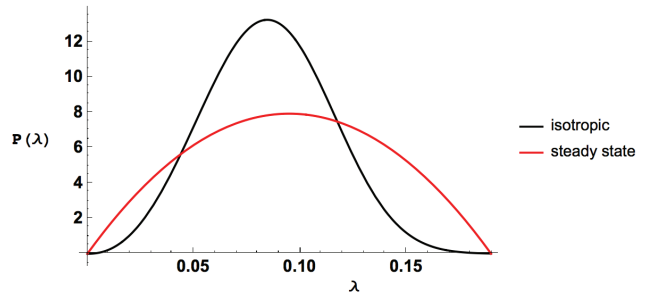}
\caption{Plots of the isotropic Weibull distribution and the Steady State 
distribution assuming that the new eigenvalues are equally distributed between 
$\lambda=0$ and $\lambda = \lambda_{max}$. The distributions have been chosen so 
that $\langle \lambda \rangle_{iso}= \langle \lambda \rangle_{ss}$.}
\label{fig1}
\end{figure}

The same conclusion can be obtained directly by analyzing the pdf in the steady 
state plastic flow.
There both $D(\gamma)$ and $ j_{in}(\lambda,\gamma )$ must become $\gamma$ 
independent, and we will assume a
constant reinjection rate $ j_{in}(\lambda) = j_{in}$ between $0< \lambda < 
\lambda_{max}$ in line with the avalanching process that occurs in the dynamics, 
while $ j_{in}(\lambda) = 0$ otherwise.
We need to solve
\begin{equation}
\label{fokkerplanck3}
0 = a_{ss} \partial [ P/\lambda ]/\partial \lambda + D_{ss}\partial^2 P/\partial 
\lambda^2 + j_{in}(\lambda ) .
\end{equation}
together with Eq.~(\ref{pad}). The solution is given by
\begin{equation}
\label{pss2}
P_{ss}(\lambda ) =  \frac{6}{\lambda_{max}}[(\lambda/\lambda_{max} ) -  
(\lambda/\lambda_{max} )^2]
\end{equation}
as a quick substitution will confirm. It is properly normalized and $\langle 
\lambda \rangle = \lambda_{max}/2$. Choosing the same value for  $\langle 
\lambda \rangle$ for both the isotropic and steady state distributions yields 
Fig.~\ref{fig1}.
%%%%%%%%%%%%%%%%%%%%%%%%%%%%%%%%%%%%%%%%%%%
%%%%%%%%%%%%%%%%%%%%%%%%%%
In addition, for this solution the probability flux $J$ at the adsorbing 
boundary and the required reinjection rate $j_{in}$ are
respectively
\begin{eqnarray}
\label{fluxess}
J  & = & \frac{-6 (a_{ss} + D_{ss})}{\lambda_{max}^2} \nonumber \\
j_{in} & = & \frac{6 (a_{ss} + 2D_{ss})}{\lambda_{max}^3} .
\end{eqnarray}
The important thing to notice is that the conclusion Eq.~(\ref{lamtozero}) is 
valid also in
the steady state, i.e. that $\theta=1$ and $\eta=0$.

We thus conclude that there exists a discontinuous transition at $\gamma=0^+$ 
such that the
Weibull distribution is swept in favor of a distribution that can support a 
finite flux at
$\lambda=0$. For completeness we discuss the evolution of the pdf upon straining 
in the Appendix.

\section{Conclusions}
\label{conclude}

\noindent Our analysis of a generic 2-dimensional system indicated the existence 
of three types of solutions:

(i)At $\gamma=0$ an isotropic Weibull solution
\begin{equation}
\label{weibull}
P_w(\lambda )  = \frac{(1 + \theta)}{\langle \lambda \rangle} (\lambda/\langle 
\lambda \rangle)^{\theta} \exp{-(\lambda/\langle \lambda \rangle)^{1 + \theta}}
 \end{equation}
 exists which is singular as $\lambda\to 0$ and for which there is no current; 
$J=0$ as $\theta \approx 2.2$
 in the case of slow quench. In
 this state of mechanical equilibrium both the scaling law (\ref{stat}) {\bf 
and} the scaling law
(\ref{etatheta}) are obeyed.

(ii) For any strain not equal to zero the statistics is fundamentally different. 
For any value of the strain, both before and after the plastic yield there 
exists a finite flux of eigenvalues towards $\lambda=0$ where there
is an absorbing boundary condition. This requires $\theta=1$. Both the scaling 
laws (\ref{stat}) and (\ref{etatheta}) appear to hold as long as 
$\gamma<\gamma_{_{\rm Y}}$. Consequently we expect to find
$\eta=0$ in agreement with the numerical simulations. If the scaling law 
(\ref{stat}) is
still obeyed, we expect to find $\beta=-1$. The direct
numerical simulation found $\beta\approx -0.92$ (cf. Fig.~\ref{simbeta}) middle 
panel. We ascribe this and the
smooth down fall to this value in Fig.~\ref{mislead} to finite size effects.

\begin{figure}
\hskip -0.5 cm
\includegraphics[scale = 0.35]{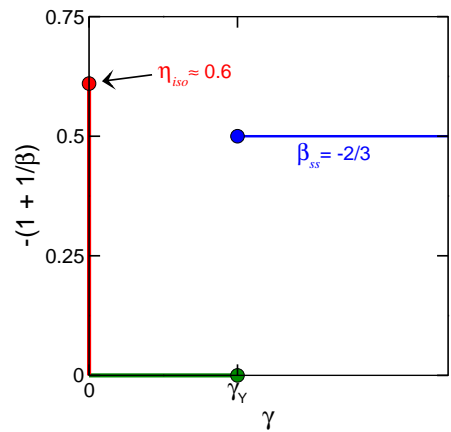}
\caption{Schematic presentation of the theoretical prediction for the $\gamma$ 
dependence of
$-(1+1/\beta)$. We reiterate that the value of this exponent at $\gamma=0$ is 
{\em not} universal,
whereas in the steady state it is universal}
\label{final}
\end{figure}

(iii) Contrary to the analysis shown in the appendix of the breakdown of the 
Weibull distribution, we do not
have a similar analysis of the yielding transition.
It was, however, possible to solve for the steady state pdf in the form of 
Eq.(\ref{pss2}).
This solution will be valid for $\gamma > \gamma_Y$. Indeed here $\theta=1$ 
(compared to $\theta = 2.2$ in the isotropic case) and there is a steady state 
flux through the system. The actual shape of $P_{ss}(\lambda)$ is not universal  
for large $\lambda$ and will depend on the form of the injection feeding 
$j_{in}(\lambda )$. In Fig.~\ref{fig1} we have assumed constant feeding for all 
$\lambda < \lambda_{max}= 2 \langle \lambda \rangle $ . In contrast the shape 
for small $\lambda$ is expected to be universal. Since $\theta=1$ we expect and 
find
in this range $\eta=0$ due to the ubiquitous scaling law (\ref{etatheta}). On 
the other hand the
scaling law (\ref{stat}) does {\bf not} hold in the flowing steady state. This 
scaling law stems from the
independence of plastic events for $\gamma<\gamma_{_{\rm Y}}$ and it fails in 
the steady state
due to the build-up of correlations between the subextensive plastic events. We 
thus expect $\beta=-2/3$
universally in this regime, in contradiction with the scaling law (\ref{stat}). 
Our expectation
for the theoretical dependence of $\beta$ on $\gamma$ is thus shown in 
Fig.~\ref{final}.
%%%%%%%%%%%%%%%%%%%%%%%%%%%%%%%%%%
%%%%%%%%%%%%%%%%%%%%%%%%%%%%%%%%%%%%%%%%%%

Finally, a word of caution is called for. As explained above in several 
occasions it is the distribution
of minimal eigenvalues of the Hessian that is the focus of our analysis. There 
is another distribution
that may be at play here, i.e. the full distribution of plastic modes. This may 
have a different
form and may be associated with other exponents. We assumed that the exponent 
$\beta$ is associated with
the distribution of minimal eigenvalues. It remains to be seen whether one can 
determine the full distribution
of plastic modes and how that distribution is related to the system-size 
dependence of $\langle \Delta \gamma \rangle$. At this point the full 
distribution of the eigenvalues of the plastic modes is not available for
a generic amorphous solid. It will be certainly worthwhile to find new methods 
of determining it.

\acknowledgments
This work had been supported in part by an ERC ``ideas" grant STANPAS. We thank 
Edan Lerner and Matthieu Wyart
for useful discussions.

\appendix

\section{The Evolution of the Probability Distribution Below Yield}

\noindent In order to study qualitatively the evolution of the probability 
distribution below yield we shall examine the simplest Fokker-Planck equation 
consistent with a singular drift velocity due to the saddle node dynamics under 
strain, namely
\begin{equation}
\label{fokkerplanck2app}
\partial P/\partial \gamma \approx a \partial [ P/\lambda ]/\partial \lambda + 
D(\gamma)\partial^2 P/\partial \lambda^2 + j_{in}(\lambda;\gamma ) \,
\end{equation}
and wish to study the evolution from the initial condition which is the 
isotropic Weibull distribution valid at $\gamma = 0$ and given by 
Eq.~(\ref{weibull}.
 Thus
 \begin{equation}
\label{pinitial}
P(\lambda;\gamma =0) = P_w(\lambda ).
\end{equation}

\noindent The easiest way to do this is to define the linear Liouville operator
\begin{equation}
L = a \partial [ 1/\lambda ]/\partial \lambda + D\partial^2 /\partial \lambda^2
\end{equation}
in terms of which we can rewrite Eq.~(\ref{fokkerplanck2app}) as
\begin{equation}
\label{fokkerplanck4}
\partial P/\partial \gamma = LP + j_{in}(\lambda;\gamma ) .
\end{equation}
This equation has the general solution $P=P_{hom}+P_{inhom}$ where
\begin{eqnarray}
\label{fokkerplanck5a}
P_{hom}(\lambda;\gamma ) & = & e^{\gamma L}P_w(\lambda ) \nonumber \\
P_{inhom}(\lambda;\gamma ) & = & e^{\gamma L}\int_{0}^{\gamma}e^{-\gamma' L} 
j_{in}(\lambda;\gamma' ) d \gamma' .
\end{eqnarray}
The inhomogeneous solution depends on the structure of the injection rate  
$j_{in}(\lambda, \gamma )$ and is not universal. The only thing we really know 
is that  the reinjection rate must ensure conservation of probability. We shall 
take the constant reinjection rate $ j_{in}(\lambda,\gamma ) = j_{in}(\gamma )$ 
between $0< \lambda < \lambda_{max}$ in line with the avalanching process that 
occurs in the dynamics, while $ j_{in}(\lambda,\gamma ) = 0$ otherwise, and 
consequently
\begin{equation}
\label{fokkerplanck5aa}
P_{inhom}(\gamma )  =  \int_{0}^{\gamma} j_{in}(\gamma' ) d \gamma' ,
\end{equation}
where $j_{in}(\gamma' )$ is fixed by the requirement that probability is 
conserved. As $P(\lambda ,\gamma )=P_{hom}(\lambda;\gamma )+P_{inhom}(\gamma )$, 
we see that on integrating this expression between $0<\lambda<\lambda_{max}$ 
that
\begin{equation}
\label{fokkerplancknorm}
1 = \int_{\sqrt{2 a \gamma}}^{\lambda_{max}} P_{hom}(\lambda;\gamma )d\lambda + 
\lambda_{max}P_{inhom}(\gamma )d \lambda ,
\end{equation}
and this yields an expression for $j_{in}(\gamma )$
\begin{equation}
\label{fokkerplancknorm2}
 j_{in}(\gamma ) = \frac{-1}{\lambda_{max}} d\large[ \int_{\sqrt{2 a 
\gamma}}^{\lambda_{max}}P_{hom}(\lambda; \gamma )d\lambda \large]/d\gamma  .
\end{equation}
Note that the lower bound of the integral involving $P_{hom}$ only extends to 
$\sqrt{2 a \gamma}$. This (as we will see below) is because $P_{hom}(\lambda < 
\sqrt{2 a \gamma};\gamma )=0$.

The homogeneous solution shows the evolution of the Weibull distribution under 
straining it and obeys
\begin{equation}
\label{fokkerplanck5}
P_{hom}(\lambda ;\gamma ) = e^{\gamma L}P_w(\lambda ) = \int_0^{\infty}dx P_w(x) 
e^{\gamma L}\delta (x - \lambda ),
\end{equation}
The Liouville operator acting on the delta function $e^{\gamma L}\delta 
(x-\lambda )= \frac{1}{\sqrt{4 \pi D \gamma }} 
\exp[{-\frac{(x-\tilde{\lambda}(\lambda;\gamma))^2}{4 D \gamma}}]$ will transform 
it into a Gaussian. In Eq.~(\ref{fokkerplanck5}) the expression 
$\tilde{\lambda}(\lambda ; \gamma)=\sqrt{\lambda^2 - 2 a \gamma}$ is the value 
that an eigenvalue $\lambda$ assumes due to drift alone. Thus we can rewrite 
Eq.~(\ref{fokkerplanck5}) as
\begin{equation}
\label{fokkerplanck6}
P_{hom}(\lambda;\gamma )  = \frac{1}{\sqrt{4 \pi D \gamma }} \int_0^{\infty}dx 
P_w(x) \exp\left[{-\frac{(x-\tilde{\lambda}(\lambda; \gamma))^2}{4 D \gamma}}\right].
\end{equation}
Using the new variable $t$ where $x = \tilde{\lambda}(\lambda;\gamma)+\sqrt{4 D 
\gamma}t$ we can rewrite Eq.~(\ref{fokkerplanck6}) as
\begin{equation}
\label{fokkerplanck7}
P_{hom}(\lambda;\gamma ) =  \frac{1}{\sqrt{\pi}} 
\int_{\frac{-\tilde{\lambda}(\lambda; \gamma)}{\sqrt{4 D \gamma}}}^{\infty}dt 
P_w(\tilde{\lambda}(\lambda ; \gamma)+\sqrt{4 D \gamma}t) \exp\left[{-t^2}\right].
\end{equation}
From Eq.~(\ref{fokkerplanck7}) we see that as $\lambda \rightarrow 
\sqrt{2a\gamma}$ from above we find
\begin{equation}
\label{fokkerplanck77}
P_{hom}(\sqrt{2a \gamma};\gamma ) =  \frac{1}{\sqrt{\pi}} \int_{0}^{\infty}dt 
P_w(\tilde{\lambda}(\sqrt{4 D \gamma}t) \exp\left[{-t^2}\right],
\end{equation}
because here $\tilde{\lambda}=0$. But for $\lambda < \sqrt{2a\gamma}$ there is 
no solution. Any nonzero value for $P(\lambda <\sqrt{2a \gamma} ,\gamma )$ must 
come from feeding combined with flow from $\lambda = \sqrt{2 a \gamma}$ and 
adsorption at $\lambda = 0$ resulting in the boundary condition $P(\lambda = 0 
,\gamma )=0$.

\noindent On the other hand if  $\lambda \gg \sqrt{2a\gamma}$, the Weibull term 
$P_w(\tilde{\lambda}(\lambda , \gamma)+\sqrt{4 D \gamma}t)$ can be expanded in 
powers of $t$ inside the integral representation for $P_{hom}$ given by 
Eq.~(\ref{fokkerplanck7}) as the exponential $\exp{-t^2}$ will ensure that only 
values of $t\ll 1$ are important. In this manner an expansion in terms of the 
gradients of the Weibull distribution can be derived. To second order we find
\begin{eqnarray}
\label{fokkerplanck8}
P_{out}(\lambda;\gamma ) & = &  C_1(\lambda;\gamma 
)P_w(\tilde{\lambda}(\lambda;\gamma)) \nonumber \\
&+ &  C_2(\lambda ,\gamma )P_w(\tilde{\lambda} (\lambda ;\gamma))' \nonumber \\
&+ & C_3(\lambda;\gamma ) P_w(\tilde{\lambda}(\lambda;\gamma))'' ,
\end{eqnarray}
where
\begin{eqnarray}
\label{fokkerplanck8b}
C_1(\lambda;\gamma ) &= & [1 +  Erf(\tilde{\lambda}(\lambda;\gamma )/\sqrt{4 D 
\gamma})]/2 \nonumber \\
C_2(\lambda;\gamma ) & =  & \sqrt{D\gamma/\pi} 
\exp{[-\tilde{\lambda}(\lambda;\gamma)^2/(4 D \gamma)]} \nonumber \\
C_3(\lambda;\gamma ) & = & (D \gamma/2)[1 - \nonumber \\
&& \tilde{\lambda}(\lambda;\gamma)/\sqrt{(\pi D \gamma)} 
\exp{(-\tilde{\lambda}(\lambda;\gamma)^2/4 D \gamma)} \nonumber \\
& + & Erf(\tilde{\lambda}(\lambda;\gamma)/\sqrt{4 D \gamma})] ,
\end{eqnarray}
and the error function is given by $Erf(x) = (2/\sqrt{\pi})\int_0^x \exp[{- t^2}] 
dt$.
Note once again that an examination of Eq.~\ref{fokkerplanck8} shows that this 
solution becomes singular when $\tilde{\lambda}(\lambda;\gamma)=0$ or for 
$\lambda < \sqrt{2 a \gamma}$. In other words this solution is also only valid 
for $\lambda > \sqrt{2 a \gamma}$. Thus for any $\gamma \neq 0$ a growing region 
exists where the Weibull solution breaks down. It is for this reason that we 
have written $P_{hom}=P_{out}$ in Eq.~(\ref{fokkerplanck8}) to represent the 
outer solution of the distribution.

\noindent We note there appears to exist  a growing boundary layer with an inner 
solution $P_{in}(\lambda, \gamma )$ for $\lambda < \sqrt{2 a\gamma}$.  To solve 
for this inner solution we write
\begin{equation}
\label{inner}
P_{in}(\lambda;\gamma) = \Sigma_n A_n(\gamma) \lambda^n \,
\end{equation}
and substitute in Eq.~(\ref{fokkerplanck2app}). For $\lambda <  \sqrt{2 
a\gamma}$, the Weibull distribution is swept away to to be replaced by 
essentially a constant flux distribution $P_{in}(\lambda;\gamma)\approx 
A_1(\gamma) \lambda$ for small $\lambda$. In that case, by matching solutions at 
$\lambda = \sqrt{2a\gamma}$ we would find $A_1(\gamma )\approx P_{out}( 
\sqrt{2a\gamma}, \gamma )/\sqrt{2a\gamma}$ due to fitting of inner and outer 
solutions. Together these arguments yield
\begin{equation}
\label{fpinner}
P_{in}(\lambda;\gamma)\approx \left[P_{out}( \sqrt{2a\gamma}, \gamma 
)/\sqrt{2a\gamma}\right] \lambda .
\end{equation}
As we can approximate Eq.~(\ref{fokkerplanck77}) by its saddle point 
approximation for small $\gamma$ we also find
\begin{equation}
\label{fokkerplanck777}
P_{hom}(\sqrt{2a \gamma};\gamma ) \approx \left[ \frac{(2 \theta D \gamma)}{e \langle 
\lambda \rangle^2}\right]^{\theta/2}\sqrt{ \frac{4 \pi D \gamma}{\langle \lambda 
\rangle(1+\langle \lambda \rangle)}}.
\end{equation}
We are now in a position to calculate the flux $J(\gamma)$ and find that at 
small strain it grows in a singular manner as
\begin{equation}
\label{fokkerplanck7777}
J(\gamma ) \sim \gamma^{\theta/2} .
\end{equation}

\end{document}